\documentclass[aps,prl,twocolumn,color,psfig,epsf,superscriptaddress]{revtex4-2}
\usepackage[english]{babel}
\usepackage[T1]{fontenc}
\usepackage{tikz}
\usepackage{amssymb,amsmath,amsfonts}
\usepackage{textcomp}
\usepackage{braket}
\usepackage{nicematrix}  
\usepackage{graphicx,color}
\usepackage[dvipsnames]{xcolor}
\usepackage[utf8x]{inputenc}
\usepackage{bm}
\usepackage{mathbbol}
\usepackage{float}
\usepackage{comment}
\usepackage[linkcolor = blue, citecolor = red, urlcolor = blue, colorlinks = true]{hyperref}

\usepackage{tabularx}
  
\begin{document}

\title{Arrested phase separation and amorphous patterns in charged binary fluids}
\author{Logan~H.~Grant}
\affiliation{SUPA, School of Physics and Astronomy, University of Edinburgh, Peter Guthrie Tait Road, Edinburgh, EH9 3FD, UK}
\author{Ana Bensabat}
\affiliation{SUPA, School of Physics and Astronomy, University of Edinburgh, Peter Guthrie Tait Road, Edinburgh, EH9 3FD, UK}
%\author{E. Pospíšilová}
\author{Eva Posp\'{i}\v{s}ilov\'{a}}
\affiliation{Institute of Materials and Machine Mechanics, D{\'u}bravsk{\'a} cesta 9, 845 13 Bratislava, The Slovak Republic}
\author{Joost de Graaf}
\affiliation{Institute for Theoretical Physics, Center for Extreme Matter and Emergent Phenomena, Utrecht University, Princetonplein 5, 3584 CC Utrecht, The Netherlands}
\author{Alexander Morozov}
\affiliation{SUPA, School of Physics and Astronomy, University of Edinburgh, Peter Guthrie Tait Road, Edinburgh, EH9 3FD, UK}
\author{Giuseppe Negro}
\affiliation{SUPA, School of Physics and Astronomy, University of Edinburgh, Peter Guthrie Tait Road, Edinburgh, EH9 3FD, UK}
\author{Ciprian~G. Pruteanu}
\affiliation{SUPA, School of Physics and Astronomy, University of Edinburgh, Peter Guthrie Tait Road, Edinburgh, EH9 3FD, UK}
\author{Tyler~N. Shendruk}
\affiliation{SUPA, School of Physics and Astronomy, University of Edinburgh, Peter Guthrie Tait Road, Edinburgh, EH9 3FD, UK}
\author{Davide Marenduzzo}
\affiliation{SUPA, School of Physics and Astronomy, University of Edinburgh, Peter Guthrie Tait Road, Edinburgh, EH9 3FD, UK}

%\affiliation{SUPA, School of Physics and Astronomy, Edinburgh EH9 3FD, Scotland, UK}
\begin{abstract}
Frustration generated by competing interactions can stabilise patterns with tunable lengthscales which evade indefinite coarsening and macroscopic phase separation. 
Here we investigate a charged binary fluid in which the tendency toward phase separation is opposed by screened electrostatic repulsion. Combining analytics and numerical simulations, we identify a regime of arrested phase separation characterised by finite-wavelength instability.
The resulting states include droplet and labyrinthine phases in two dimensions, and droplets, concentric shells and tubular structures in three dimensions. Surprisingly, some of these structures are amorphous morphologies that emerge reproducibly from different initial conditions. Free-energy estimates suggest that these disordered structures may be  thermodynamically stabilised by frustration, %or within $k_BT$ of a thermodynamic minimum, so they can be viewed as a proper phase 
rather than being kinetically trapped. Our results identify a simple mechanism through which the competition between phase separation and screened electrostatics can generate equilibrium amorphous microphases with glassy features. 
\end{abstract}

\maketitle

Frustration arising from competing interactions is a generic mechanism for the emergence of complex and tunable spatial patterns in physical systems~\cite{seul1995,andelman2009}. A paradigm model for such a phenomenon is provided by the competition between short-range attraction and long-range repulsion (SALR), which can generate modulated phases and patterns including lamellae, stripes, bubbles and labyrinthine structures~\cite{jackson1994,haw2010,low1994,schmalian2000,liu2019,spivak2004,stradner2004,campbell2005,chakrabarti2011,ortix2006}. Such behaviour has been studied in magnetic systems, dipolar fluids, block copolymers and frustrated spin models, where the system is unable to simultaneously minimise all local energetic contributions~\cite{jackson1994,vannimenus1977}. More generally, frustration is a recurring organising principle in condensed matter physics, often associated with complex free-energy landscapes and slow relaxation, such as in the physics of glasses~\cite{zaccarelli2007,tarjus2005,sciortino2004,kivelson1995}. 

Here we consider a specific example of frustration, that arises in a charged binary mixture from the competition between a baseline intermolecular attraction which provides the thermodynamic drive to demix, and screened electrostatic repulsion which favours mixing. This system is relevant, for instance, to the physics of polyelectrolyte phase separation and of self-attractive suspensions of charged colloids~\cite{boroudjerdi2005,dean2014,zheng2021}, where electrostatic screening is controlled through salt concentration. As we discuss below, it may also be of relevance to understand the formation of protein-nucleic acid condensates in living cells~\cite{shin2017}.  
As expected~\cite{seul1995}, this competition results in arrested separation: rather than coarsening indefinitely, the system evolves towards finite-sized domains whose characteristic scale is selected by the balance of interfacial and electrostatic energies. In our case, this can be captured by suitable electrocapillary numbers. In line with our numerical results, linear stability analysis reveals an intermediate regime between the linearly stable and unstable regions. In this intermediate regime, long-wavelength coarsening is suppressed while finite-wavelength modes remain unstable, signalling the arrest of phase separation.

The main result of our work is that this frustrated regime gives rise not only to regular modulated structures, such as, in two dimensions, hexagonal crystals of droplets or lamellar phases as in Brazovskii-like theories~\cite{brazovskii1975,brazovskii1987,elder2002}, but it also yields a panoply of amorphous morphologies. These amorphous morphologies emerge reproducibly across independent initial conditions. In two dimensions, they include disordered labyrinthine and droplet-like states. In three dimensions we observe structures with coexisting concentric shells and meandering tubes, reminiscent for instance of the Golgi apparatus inside the cytoplasm of living cells~\cite{lowe2011}. Although lacking obvious crystalline order, these morphologies appear consistently with different types of initial conditions and persist over long times, suggesting that they may correspond to thermodynamically stable disordered states rather than simply long-lived metastable transients. In other words, these structures may share some of the features of equilibrium glasses~\cite{zaccarelli2007}.

%\section{Equations of Motion}
We consider a binary mixture made up by a charged component and a neutral one. We call the local density of the charged component $\phi$, and the electrostatic potential $\psi$. The system is described by the free-energy density
\begin{equation}
    f = \frac{a}{4}\phi^2\left(\phi-1\right)^2 +\frac{k}{2}\left(\nabla\phi\right)^2-\frac{\varepsilon}{2}\left(\nabla\psi\right)^2-\frac{\varepsilon}{2\lambda_d^2}\psi^2+c\phi\psi.
    \label{freeenergydensity}
\end{equation}
In Eq.~(\ref{freeenergydensity}), the first term on the right hand side with $a>0$ favours phase separation, the second term penalises interface formation, with $k$ determining simultaneously interfacial width and surface tension, $\varepsilon$ is the dielectric permittivity, which we assume is spatially constant, $\lambda_d$ is the Debye screening length of the medium, and $c$ is a coupling constant which can be thought of as proportional to the charge density of the charged component. %of the component described by $\phi$. 

The dynamics of the system are governed by two coupled equations: a Cahn-Hilliard-like (model B) equation for the globally conserved density $\phi$, and a Poisson-Boltzmann equation for the electrostatic potential $\psi$, respectively written in dimensionless form as
\begin{eqnarray}
    \frac{\partial\phi}{\partial t} & = & \nabla^2\left[\phi\left(\phi-1\right)\left(\phi-\frac{1}{2}\right)-\nabla^2\phi+\psi\right],
    \label{eq:non-dim-phi-psi} \\ \nonumber
    0 & = & \nabla^2\psi-a_1\psi+a_2\phi,
%    \label{eq:non-dim-screening-psi}
\end{eqnarray}
where we used $\sqrt{k/a}$, $Ma^2/k$ and $a/c$ as the scales for length, time and $\psi$ respectively. The same equations of motion hold in 2D and 3D, and we shall consider both cases in what follows. Eqs.~(\ref{eq:non-dim-phi-psi}) reveal the key dimensionless parameters determining the physics to be 
\begin{equation}
    a_1=\frac{\delta^2}{\lambda_d^2}\quad\text{and}\quad a_2=\frac{1}{a}\frac{c^2\delta^2}{\varepsilon},
\end{equation}
where $\delta=\sqrt{k/a}$ is the interfacial width. %(and the rescaled lengthscale of the system). 
Physically, $a_1$ represents the squared ratio of lengthscales within the system. The cases $a_1=0$ corresponds to unscreened Coulomb electrostatics, while $a_1\to \infty$ represents a system with no electrostatics. The quantity $a_2$ can be written as $\left(c^2k\right)/\left(a^2\varepsilon\right)$, which can be interpreted as the ratio between the typical size of the interfacial force $k$ over the electrostatic force $\delta^2\varepsilon (\nabla\psi)^2\sim \varepsilon (a/c)^2$. Hence, $a_2$ is an electrocapillary number. 

%\subsection{Linear Stability Analysis}
We begin by analysing the response of the system to small perturbations, through a linear stability analysis. We therefore consider an initial homogeneous state with a small spatially dependent random noise added to it: 
\begin{equation}
    \phi=\phi_0+\delta\phi,
    \quad %\text{where}\quad
    \delta\phi = e^{\Gamma t}e^{i{\mathbf k}\cdot{\mathbf x}}.
    \label{eq:lsa-start}
\end{equation}
The modes of fluctuations can be determined via the dispersion relation $\Gamma(k)$, where ${\mathbf k}$ is the wavevector of the perturbation. The corresponding electric potential 
%\begin{eqnarray}
    $\delta\psi=\frac{a_2}{k^2+a_1}\delta\phi$.
%\end{eqnarray}
can be found by Fourier transforming Eqs.~(\ref{eq:non-dim-phi-psi}). Discarding all non-linear $\delta\phi$ terms results in the dispersion relation
\begin{equation}
    \Gamma = -k^2\left[\hat{\phi}+k^2+\frac{a_2}{k^2+a_1}\right],
    \label{eq:dispersion-relation}
\end{equation}
where $\hat{\phi}=\frac{\partial^2f}{\partial\phi^2}\big\rvert_{\phi_0}=3\phi_0^2-3\phi_0+\frac{1}{2}$.

\begin{figure}[tb]
    \centering
    \includegraphics[width=1\linewidth]{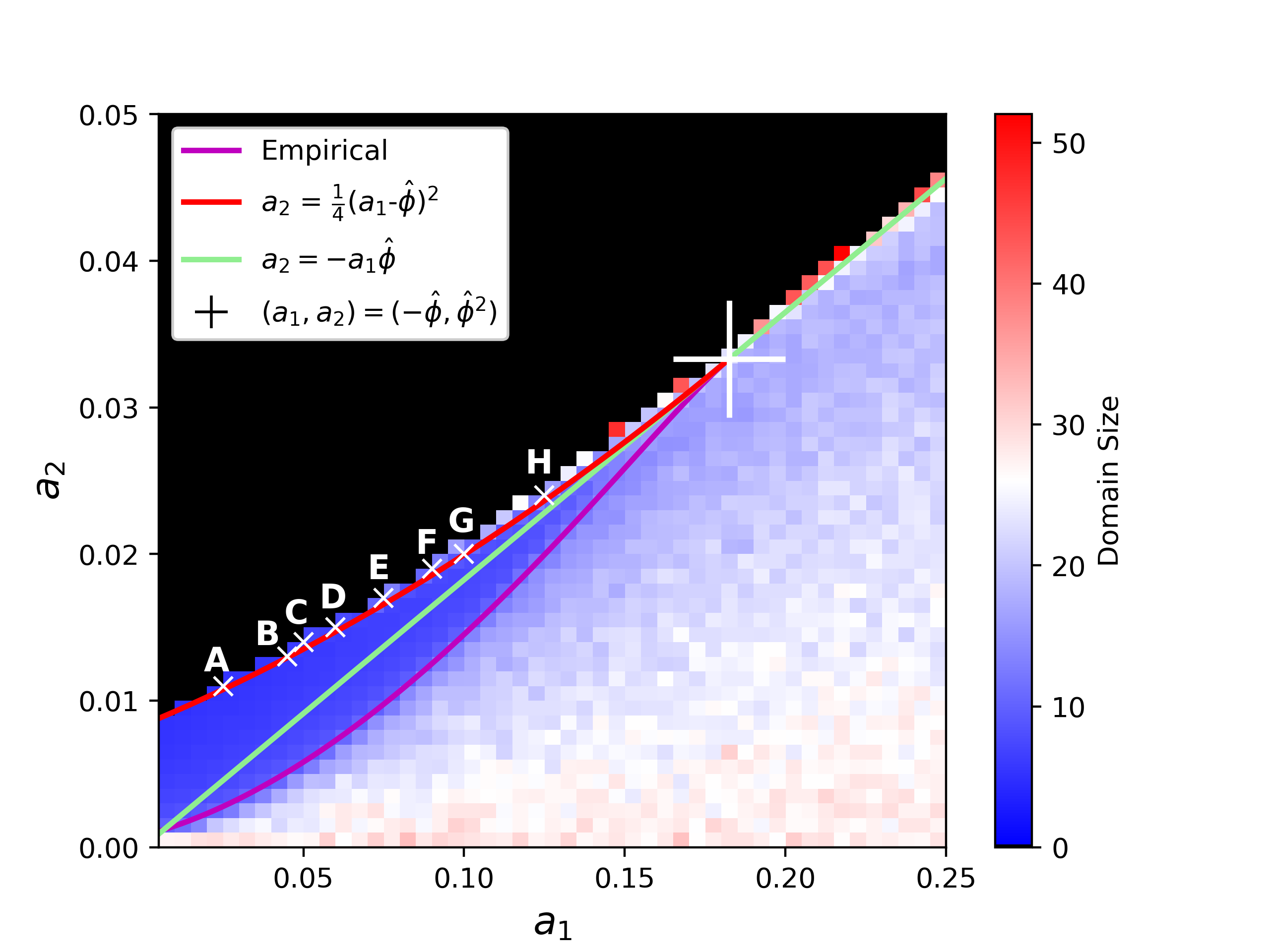}
    \caption{State diagram for $\phi_0=0.35$, where simulations start from homogeneous and a uniform noise of $\delta\phi=0.1$, with theoretical phase boundaries from linear stability analysis overlaid. Each pixel represents the domain size of a state close to equilibrium. %- this results in the three expected regions. 
The upper boundary (red) matches with the simulations, however the lower boundary (green) differs; a numerical estimate (pink) for the lower boundary is added. The point where the upper and lower boundaries merge is highlighted with a white cross. The points A-H correspond to equilibrium states shown in Figs.~\ref{fig:equilibrium-states} and~\ref{fig:equilibrium-states-suppl}.}
    \label{fig:phase-diagram-sim}
\end{figure}

Analysing the shape of $\Gamma$ produces a linear stability diagram in the parameter space of $(\phi_0,a_1,a_2)$, showing regions of linear instability and where the instability occurs for intermediate wavevectors (End Matter; Fig.~\ref{fig:lsa}). %Figure \ref{fig:lsa} and Figure \ref{fig:bin-spin}).
Three regions are identified: a linearly stable region (Fig.~\ref{fig:lsa}; black), a linearly unstable region (Fig.~\ref{fig:lsa}; blue) and an arrested separation region between the two (Fig.~\ref{fig:lsa}; red). The upper boundary between the arrested and stable regions corresponds to cases for which there is only one positive non-zero solution to $\Gamma=0$ (stationary point shown as the red line in Fig.~\ref{fig:lsa}; inset). %Disregarding the case of $k=0$ and solving for $k^2$,
We obtain
\begin{equation}
    k^2=-\frac{(\hat{\phi}+a_1)}{2}\pm\frac{1}{2}\sqrt{(a_1-\hat{\phi})^2-4a_2}.
\end{equation}
Considering the physical case in which $k$ is real, the upper boundary is given by
%\begin{equation}
$a_{2}=\frac{1}{4}(a_1-\hat{\phi})^2$
%\label{eq:upper-boundary1}
%\end{equation}
for $a_1<-\hat{\phi}$, and by
%\begin{equation}
$a_{2}=-a_1\hat{\phi}$
%\label{eq:upper-boundary2}
%\end{equation} 
for $a_1>-\hat{\phi}$ (Fig.~\ref{fig:phase-diagram-sim}; red line).

The lower boundary can be determined as the point at which the curvature of $\Gamma\left(k\right)$ at $k\to 0$ changes sign: %. It occurs when the dispersion relation switches from curving downwards to upwards,
%\begin{equation}
%    \frac{\partial^2\Gamma}{\partial k^2}=-2\left[\hat{\phi}+6k^2+\frac{a_2}{k^2+a_1}-\frac{5a_2k^2}{(k^2+a_1)^2}+\frac{4a_2k^4}{(k^2+a_1)^3}\right]
%\end{equation}
\begin{equation}
    \frac{\partial^2\Gamma}{\partial k^2}\Bigg{|}_{k=0}=\hat{\phi}+\frac{a_2}{a_1}=0\quad\Rightarrow\quad a_2=-a_1\hat{\phi} ,
\end{equation}
which is the green line in Fig.~\ref{fig:phase-diagram-sim}. 
The upper and lower boundaries intersect, indicating a triple point exists, which occurs at $(a_1,a_2)=(-\hat{\phi},\hat{\phi}^2)$. For a global composition of $\phi_0=0.35$, the triple point is at $(a_1,a_2)\simeq(0.183,0.033)$ (Fig.~\ref{fig:phase-diagram-sim}; white cross).

%\subsection{Numerical simulations and phase diagram in 2D}

\begin{figure}[tb]
    \centering
    \includegraphics[width=1\linewidth]{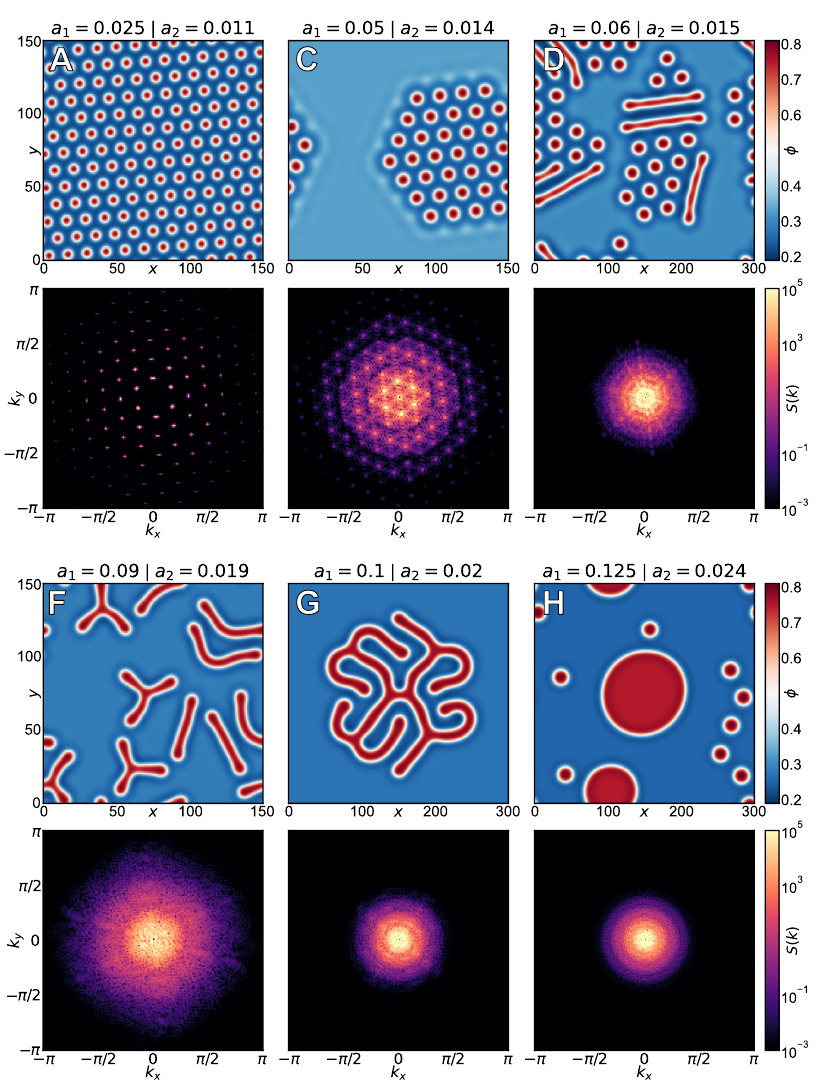}
    \caption{Equilibrium states corresponding to states A, C, D, F, G, and H as labelled in  Fig.~\ref{fig:phase-diagram-sim}.}
    \label{fig:equilibrium-states}
\end{figure}

To test these predictions, %our linear stability analysis and explore the non-linear regime of the system, 
we perform extensive numerical simulations in two dimensions governed by Eqs.~(\ref{eq:non-dim-phi-psi}); details of the numerical scheme employed to do so are given in~\cite{SI}. %under an initial spinodal decomposition protocol~\cite{SI}. 
Starting, as in the linear stability analysis, from a homogeneous state supplemented with a small, random thermal noise ($\delta\phi = 0.1$), the system evolves dynamically toward a (metastable) steady state. Depending on the selected dimensionless parameters $(a_1, a_2)$, the long-time morphologies settle into one of three distinct thermodynamic regimes: a completely mixed uniform phase, macroscopically phase-separated domains, or microphase-separated arrested structures (Figs.~\ref{fig:phase-diagram-sim}, \ref{fig:equilibrium-states}, and~\cite{SI}). 
%The microphase separated region can be quantified by average size of patterns
While the % phase 
boundary %mapping the transition 
between microphase separated patterns and the uniform mixed phase is well captured by linear stability (Fig.~\ref{fig:phase-diagram-sim}; red line), the numerically estimated boundary separating macroscopic and microphase (arrested) separation exhibits significant deviations from the linear prediction (Fig.~\ref{fig:phase-diagram-sim}; green line); this is expected as the end pattern requires knowledge of late-time dynamics inaccessible to linear stability.
%This discrepancy highlights the well-known prominent role of non-linear mode-coupling effects in determining the late-time coarsening dynamics of the system~\cite{chaikin1995}.

The morphologies of microphase separated structures include hexagonal or nearly hexagonal crystals of droplets, lamellar patterns, and combination of the two (Figs.~\ref{fig:equilibrium-states}A, C, D, and \ref{fig:equilibrium-states-suppl}). A two-mode harmonic analytical approximation accounting for the presence of modulated (microphase separated) patterns qualitatively reproduces both the overall shape of the phase diagram and the positioning of the droplet and lamellar phase within it (see End Matter, Fig.~\ref{2-mode-dia}), thereby supporting the findings of our simulations. The analytical approximation however shows that the homogeneous state is also unstable to phase separation or the formation of modulated patterns also above the upper boundary in Fig.~\ref{fig:phase-diagram-sim}, because non-linear perturbations from the isotropic phase can lead to nucleation also in systems which are linearly stable to small perturbations. 

The most notable feature of our numerical results is that, in the vicinity of the transition line between microphase separated patterns and the uniform phase, a slew of structurally disordered, amorphous morphologies emerges (Fig.~\ref{fig:equilibrium-states}). First, by systematically tracking the steady-state snapshots as a function of the energy-scale ratio $a_2$, we observe that the system progresses from a highly ordered, periodic crystal of hexagonal droplets (Fig.~\ref{fig:equilibrium-states}A) to localised crystalline clusters coexisting with the uniform background (Fig.~\ref{fig:equilibrium-states}C; Suppl. Movie 1). These localised motifs are examples of ``homoclinic snaking'' found in dynamical systems with pattern forming regimes~\cite{bergeon2008,kozyreff2006}. As frustration increases, these configurations give way to a coexistence of crystalline droplets and elongated lamellar segments (Fig.~\ref{fig:equilibrium-states}D), eventually resulting in disordered lamellar segments and sprawling labyrinthine networks (Fig.~\ref{fig:equilibrium-states}F,G; Suppl. Movie 2). In these disordered states, branching events are frequently observed, giving rise to trefoils (Fig.~\ref{fig:equilibrium-states}F) and labyrinthine patterns (Fig.~\ref{fig:equilibrium-states}G). Interestingly, these labyrinthine configurations, as well as some of the droplet morphologies~\cite{SI}, appear amorphous. This structural disorder is qualitatively confirmed by a structure factor analysis (often used to identify amorphous structures~\cite{henrich2010}), which reveals absence of discrete Bragg peaks, displaying instead diffuse rings, indicative of isotropic short-range correlation without long-range translational order as expected for amorphous patterns 
(Figs.~\ref{fig:equilibrium-states}; see \ref{fig:equilibrium-states-suppl} for structure factors of larger scale simulations).

\begin{figure}[tb]
    \centering
    \includegraphics[width=1\linewidth]{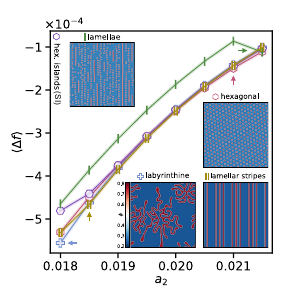}
    \caption{Plot of $\langle \Delta f\rangle$, the average free energy density difference (with respect to the homogeneous mixed state) as a function of $a_2$, for the most common morphologies at $\phi_0=0.35$ and $a_1=0.1$: regular lamellae, lamellar stripes in a uniform background, hexagonal droplet crystal, hexagonal crystal islands in a uniform background, and amorphous labyrinthine states. A difference in $f$ of $10^{-6}$ can be mapped to $1$ Pa, which integrated over a volume of $l^3$ with $l\sim 100$ nm gives $\sim 0.24$ $k_BT$~\cite{SI}. %Therefore, states with differences in $f$ less than $\sim 4\times 10^{-6}$ require including entropy in their free energy, which likely stabilises amorphous states. 
    Kinks in the curves typically corresponds to points for which final patterns differ in morphology from initial ones (Fig.~\ref{fig:extraconfigs_free_energy_vs_a2}). 
    Additional results and details are given in~\cite{SI}, together with explicit initial patterns used.
    %For a small range inside the arrested region, $a_2=[0.01825,0.01925]$, the labyrinthine networks result in the lowest free energy of the three states. 
    %Each simulation starts from an initial state (A, B, C) and is run until the free energy stabilises. %close to equilibrium. 
    %These initial states are formed from homogenous and noise ($\phi_0=0.34998$, $\delta\phi=0.01$): 
    %Initial states A and B are formed at $\phi_0=0.34998$  by quenching from the isotropic state, and with $a_2=0.01825$ and $a_2=0.01975$ respectively; C is formed at $a_2=0.02$ with a large fluctuation in the centre before the addition of noise. Circles denote the point at which each morphology is no longer stable. Regular lamellar morphologies result in higher free energies in this range. 
    }
    \label{fig:free_energy_vs_a2}
\end{figure}

To elucidate whether these amorphous patterns represent thermodynamically stable states or %long-lived, 
kinetically trapped states on the way to a more stable regular modulated phase, we evaluate the average of the free energy density %, Eq~(\ref{freeenergydensity}) 
over the whole simulation domain, as a function of  parameters (Fig.~\ref{fig:free_energy_vs_a2}, with %computational reasons we have only considered 
a fixed value of $a_1=0.1$). We compare the late-time free energy obtained by starting from several competing configurations, namely amorphous labyrinthine networks (as in Suppl. Movie 2), lamellar phases, lamellar stripes within a uniform background, hexagonal crystals, islands of hexagonal crystals within a uniform background (as in Suppl. Movie 1, see~\cite{SI}), fully demixed or completely mixed states.  While the last two cases can be treated analytically, to estimate the average $f$ for the others we evolved a system with the desired pattern dynamically to find a metastable local minimum with the corresponding morphology. 
Our key finding is that there is a parameter regime where amorphous structures 
%in which the disordered amorphous structures 
have values of $f$ that are very close to that of the most stable ordered structure (such that the difference between the corresponding free energies is $\le k_BT$ for a volume corresponding to the typical pattern size, see~\cite{SI}): therefore, entropy may stabilise the amorphous states in this regime. %for a mapping between simulation and physical units). 
Remarkably, for two of the nine parameter sets we analysed quantitatively, even the thermodynamically most favourable structure of those we probed is an amorphous one (see the ``lamellae'' and labyrinthine snapshots in Fig.~\ref{fig:free_energy_vs_a2}).
Therefore,  
our charged binary mixture may provide a 
realisation of an equilibrium glass, stabilised by competition between intermolecular attraction and screened electrostatics. %repulsion.  

Screening is an essential ingredient for amorphous patterns to arise, as, for purely Coulombic repulsion, $a_1\to 0$, we always observe %our simulations always show the formation of 
hexagonal droplet crystals or fully demixed structures. In~\cite{SI}, we verify that amorphous patterns also arise for different $\phi_0$.

%%%%%%%%%%%%%%%%%%%%%%%%%%%%%%%%%%%%%%%%%%%

\begin{figure*}[tb]
    \centering
    \includegraphics[width=1\linewidth]{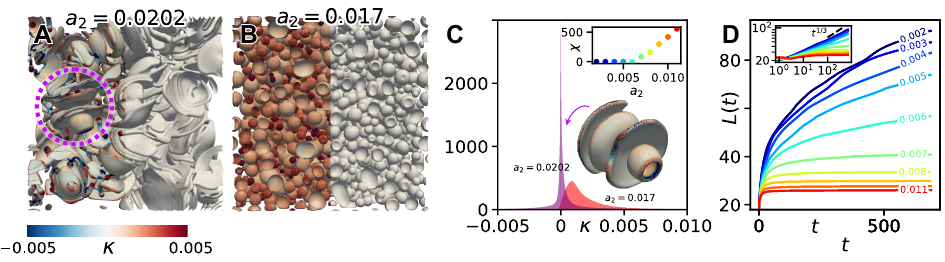}
    \caption{Morphology and topology of arrested states in 3D. (A-B) Snapshots for $a_1=0.1$. Right halves show contours of $\phi \approx0.2$ and left halves colour them according to the local Gaussian curvature $\kappa$. (C) Probability density functions (PDF) of %the local Gaussian curvature 
    $\kappa$ for the cases in A and B. The snapshot corresponds to circled structure in A. Inset: late-time Euler characteristic $\chi$ for the cases in D. (D) Average domain length $L(t)$ for $a_1=0.05$ and $a_2\in[0.002,0.011]$. Inset: Log-log scale.}
    \label{fig:fig4}
\end{figure*}

%3D results numerical simulations 
As most of the amorphous patterns we observe in 2D are labyrinthine, their surface tension cost is high, and it is unclear whether to expect that the different interfacial physics transforms them into tubular or sheet-like patterns, or even destabilises them altogether in 3D. To address this question, we extend our numerical investigations to 3D. The results show both microphase separated droplet patterns and amorphous morphologies: example snapshots are shown in Figs.~\ref{fig:fig4}A,B. The amorphous pattern shown in Fig.~\ref{fig:fig4}A shows that going to 3D significantly enriches the complexity of the ensuing morphologies. Branched lamellae which could be formed by extending labyrinthine 2D patterns %such as in Fig.~\ref{fig:equilibrium-states}F,G 
along the third dimension are not stable, and instead we observe meandering tubular structures coexisting with curved shells which can form some approximately concentric layers. As in 2D, these amorphous structures are stable close to the transition to the uniform phase. %and amorphous structures require larger values of $a_2$. 
For smaller $a_2$, we observe droplet emulsions, where the main difference with respect to their 2D counterpart is that they are usually polydisperse at the timescales which we can reach. %numerically. 

Quantitatively, we see that the amorphous regime, composed of shells and tubular filaments, is characterised by both negative and positive local curvature (Fig.~\ref{fig:fig4}A,C) with an average curvature near $\kappa\simeq 0$. On the other hand, the self-organised droplet suspension has positive local curvature as expected from their spherical geometries. 
%growth law and Euler characteristic
The transition between phase separation and droplets is apparent from the late-time Euler characteristic $\chi$~\cite{SI} (Fig.~\ref{fig:fig4}C, inset): $\chi$ remains near zero in the macrophase separated regime but then rises sharply with $a_2$ as droplets form. Correspondingly, the characteristic domain length follows Lifshitz-Slyozov growth $L(t)\sim t^{1/3}$~\cite{chaikin1995} for small electrocapillary numbers ($a_2\to 0$). As $a_2$ increases, domain growth instead flattens completely, confirming microphase separation with domain size selection (Fig.~\ref{fig:fig4}D; inset). 
It is notable that the exotic 3D amorphous structures we observe share qualitative features with the endoplasmic reticulum and Golgi apparatus which self assemble into the cytoplasm of eukaryotic cells~\cite{lowe2011}, suggesting that these highly complex organelles may arise from an underlying frustration mechanism.

In summary, we have investigated the phase-separation dynamics of a charged binary fluid, and showed that the intrinsic competition between a thermodynamic driving force to demix and a screened electrostatic repulsion acts as a physical mechanism leading to structural frustration. This competition prevents macroscopic phase separation, %and selects a finite wavelength for domain growth, 
inducing an arrested phase-separation regime in line with the well-known short-range attraction and long-range repulsion mechanism. %While linear stability analysis accurately predicts the primary onset of modulated patterns from the homogeneous phase, the deep non-linear regime reveals a richer morphological landscape. 
Our main finding is the discovery of amorphous, labyrinthine networks and localised coexistent structures in two dimensions, near the boundary between the patterned microphases and the uniform mixed state. Our simulations suggest that these disordered configurations, once entropy is accounted for, may be thermodynamically more stable than competing periodic structures, such as regular lamellae or hexagonal droplet crystals. Additionally, our simulations suggest that the basin of attraction of amorphous states is significantly larger. Our system may therefore provide a surprisingly simple example of an equilibrium structural glass. Much like the amorphous liquid crystalline ``blue fog''~\cite{henrich2011,wright1989,carenza2022}, where the proliferation of disclination lines driven by chiral frustration suppresses long-range crystalline order, the amorphous networks observed here are fundamentally driven by frustration due to electrostatic effects. %In the present case, frustration arises due to electrostatic effects. 
Finally, 3D simulations confirm the generation of amorphous structures, whilst simultaneously uncovering highly complex topologies completely absent in lower dimensions, such as concentric multi-layered shells coexisting with tubular filaments and vesicular geometries.
In the future, it would be of interest to study the effect of hydrodynamic interactions~\cite{capuani2004,giupponi2011} on the self-assembly dynamics of amorphous patterns.

Beyond applicability to charged binary systems~\cite{jiang2019}, we suggest that our results may be, perhaps surprisingly, qualitatively relevant to the biophysics of intracellular patterns in living cells. 
%organelles 
The competition between intermolecular attraction and screened electrostatics, and the resulting frustration, provides an interesting and simple pathway to arrest liquid-liquid phase separation inside cells to form protein-nucleic acids condensates of self-selected sizes~\cite{shin2017,sing2020,dutagaci2021,brackley2017}. %Furthermore, the amorphous structures we find in 3D are reminiscent of intracellular organelles, such as the endoplasmic reticulum and the Golgi apparatus~\cite{lowe2011}, which can be modelled as networks of curved shells coexisting with tubular structures and droplets. 
Electrostatics may be expected to play a minor role in the cytoplasm of cells, since the Debye length is about $1$ nm under physiological salt concentration. However, our results show that amorphous structures are prevalent around the triple point, which occurs in the presence of screening for finite values of $a_1$. This can be realistic for intracellular condensates, for which the interfacial width may be of molecular size and hence smaller than or comparable to the Debye length.  
%Additionally, in the context of non-equilibrium physics, a natural next step is to explore how these equilibrium amorphous phases respond to external driving forces, such as steady or oscillatory shear flow.

%%%%%%%%
%\section*{Acknowledgments}
\paragraph{Acknowledgments.}
\label{acknowledgements}
%%%%%%%%
E. Posp{\'i}\v{s}ilov{\'a} acknowledges support from the National Scholarship Programme of the Slovak Republic for the Support of Mobility of Students, PhD Students, University Teachers, Researchers and Artists. The programme is funded by the Ministry of Education, Research, Development and Youth of the Slovak Republic. %The two-mode phase diagrams were obtained using the computational resources procured in the national project 
We thank support from the National Competence Centre for High Performance Computing (project code: 311070AKF2) funded by European Regional Development Fund, EU Structural Funds Informatization of society, Operational Program Integrated Infrastructure.

For the purpose of open access, the authors have applied a Creative Commons Attribution (CC BY) licence to any Author Accepted Manuscript version arising from this submission.

\clearpage

\appendix
\setcounter{equation}{0}
\renewcommand{\theequation}{A\arabic{equation}}

\section{END MATTER}

\subsection{Linear stability diagram}
\begin{figure}[tb]
    \centering
    \includegraphics[width=1\linewidth]{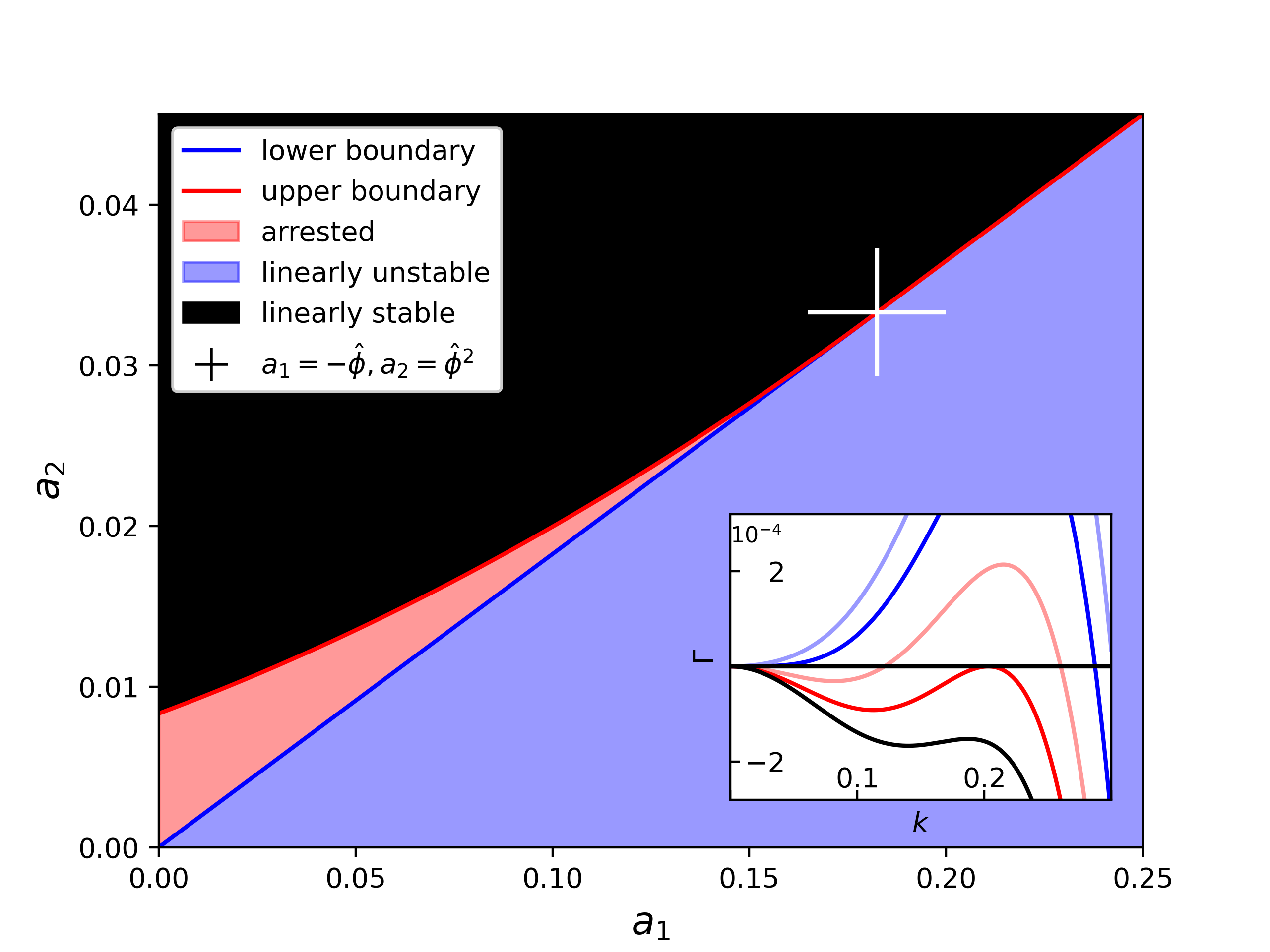}
    \vspace*{-0.25cm}
    \caption{The linear stability diagram in the $(a_1,a_2)$ plane for $\phi_0=0.35$, showing the three regions of how the system is expected to respond to small fluctuations; stable, unstable and arrested. The boundaries are determined using the dispersion relation, $\Gamma$, which is displayed in the smaller diagram. Enclosed area both above and below $\Gamma=0$ results in arrested separation; if $\Gamma<0$ for all $k$ then the system is stable towards linear perturbations, and if the enclosed area only exists above then the system is unstable. Examples of $\Gamma$ are given for each region (black, pale red, pale blue) alongside the boundaries (red, blue). The point where the upper and lower boundaries merge is highlighted with a white cross.}
    \label{fig:lsa}
\end{figure}

Figure~\ref{fig:lsa} shows the linear stability diagram, which predicts the relative stability of the homogeneous state with respect to a small noise. When the homogeneous phase is unstable, we distinguish between cases where the instability extends to $k\to 0$, and hence to a wavelength corresponding to the whole system, and cases where the instability only exists in a finite range of wavevector not extending down to $k\to 0$, which we associate with pattern formation and arrested phase separation. The two boundaries are compared with the state diagram resulting from simulations of the full non-linear model initialised in the same way in Fig.~\ref{fig:phase-diagram-sim}.

% \vspace*{-0.75cm}
\subsection{Phase Diagram in the Two-Mode Approximation}

%\subsection{Energy-Minimising States}
To provide theoretical support for the state diagram shown in the main text in
%amorphous labyrinthine patterns being the global minimum, as well as providing an alternative to the homogeneous-initiated phase diagram in Figure 
Fig.~\ref{fig:phase-diagram-sim}, we study the equilibrium of the model defined by Eq.~(\ref{freeenergydensity}) %Eqs.~(\ref{CH},\ref{DH}) 
by means of a two-mode approximation,~\cite{chdh}. 

%we produce analytical descriptions of the standard states: homogeneous, lamellar stripes and hexagonal droplet suspensions. We compute their free energies, as well as the phase coexistence of each pair. We also include the macroscopic bulk free energy. The modulated states are determined using amplitude and wavevector minimisations of 2-mode approximations. The phase coexistences are then determined using the lever rule.

%\noindent \textbf{Methods:} 

\paragraph{Methods.}
Four phases are considered in this analysis: (1) the {\it homogeneous} solution with %~constant charge density 
$\phi_{0} = 0.35$, denoted in the present diagrams as ``H''; (2) the macroscopic {\it charge-separated} phase ``S'', consisting of~two homogeneous liquids with~constant charge densities $\phi_{1} < \phi_{0}$ and $\phi_{2} > \phi_{0}$; %~\footnote{This differs from the bulk phase separation as found in the Cahn-Hilliard model. That is, the liquid with the concentration $\phi_{2} > \phi_{0}$ carries the net electric charge density $c$, increasing the free-energy density of the system, $\mathcal{F}/V$, through the repulsive electrostatic part.}; 
(3) the {\it lamellar} pattern ``L''; and (4) the {\it hexagonal-droplet} pattern ``D''. 

The free-energy density of the homogenous phase and charge-separated phase %, $f_H$ and $f_S$, 
can be found exactly. %and are respectively given by
%\begin{eqnarray}
%    f_H & = & \frac{1}{4}\phi^2(\phi-1)^2+frac{a_2}{2a_1}\phi^2, \\ \nonumber
%    f_S & = &\frac{a_2}{2a_1}\left(\bar{\phi}-\frac{a_2}{2a_1}\right).
%\end{eqnarray}
On the other hand, the free-energy densities of the lamellar and hexagonal-droplet phases L and D can only be found by an approximation. To estimate them, we use a two-mode harmonic approximation~\cite{2mode}. This approach describes the modulation of the charge concentration $\phi(x,\,y)$ around its mean value $\phi_{0}$ as a sum of two cosines at the fundamental and the doubled frequency. 

Both of the microphase-separated patterns L and D are resolved in~this two-mode harmonic approximation. The region of coexistence of the homogeneous (H) and droplet (D) phases is designated as ``H/D''; similarly the other regions of coexistence are denoted as ``H/L'' and ``D/L''. Coexistence lines for~the phase separation are identified via the common-tangent construction~\cite{chull}. For~the two-mode lamellar and hexagonal-droplet suspension patterns, three parameters are numerically optimised: the amplitudes $A$ and $B$ of the cosines (for all combinations of~signs) and the magnitude of~the wave vector, $k > 0$. 

This is accomplished using the \texttt{scipy.optimize} package, utilising both \texttt{minimize} and \texttt{differential\_evolution} methods. The lowest free-energy output is stored for~each pattern at every phase-space point. \texttt{differential\_evolution} is performed only once for~each of the numerically optimized patterns. The much faster and less robust \texttt{minimize} optimization scheme is run from~26 distinct initial conditions, including the output of~the numerical optimization at~the preceding phase-space point. A total of $500$ grid points along each coordinate are computed using periodic boundary conditions. 

\begin{figure}[tb]
\centering
\includegraphics[width = 1.0\linewidth]{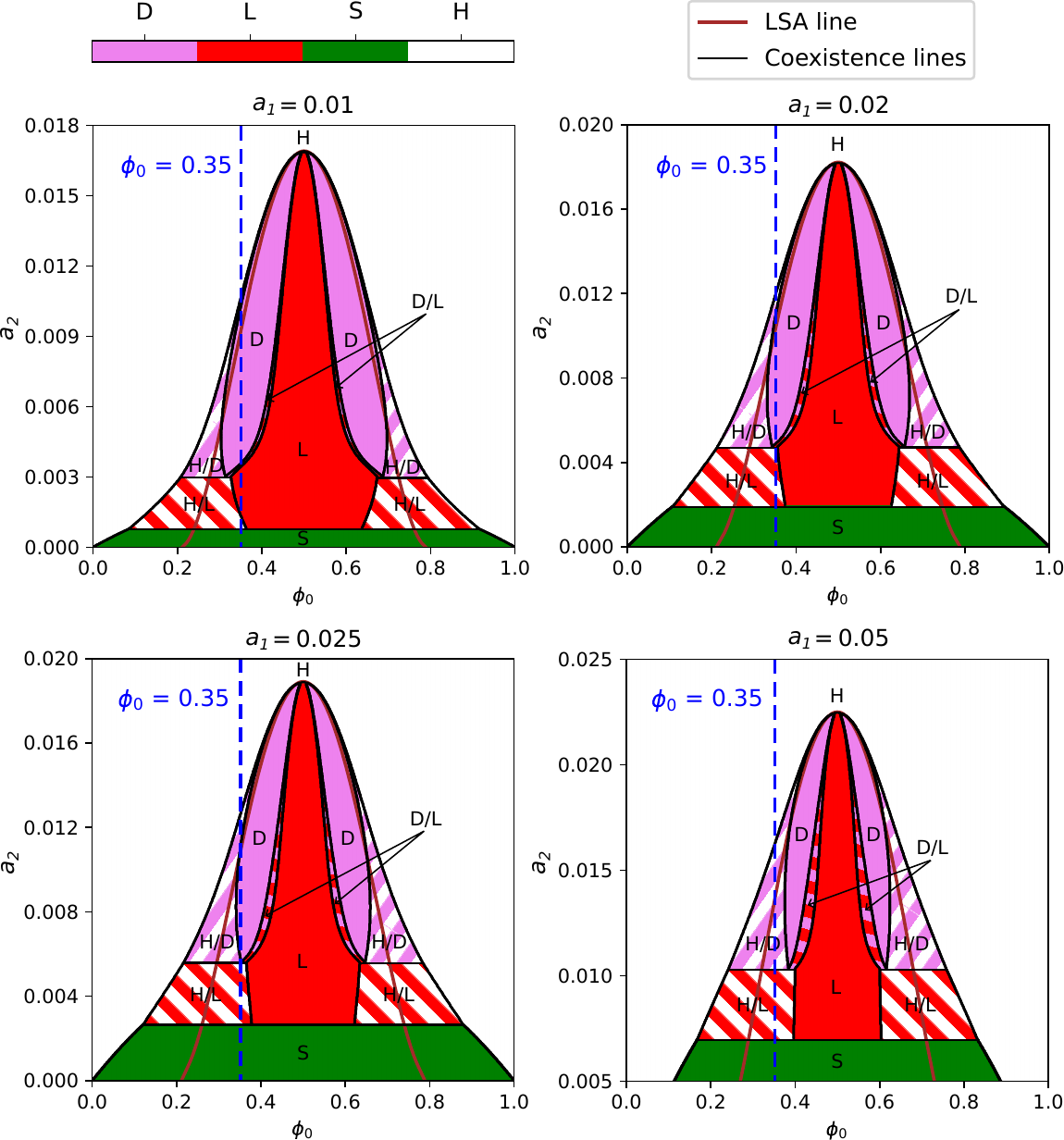}
%\vspace*{-0.25cm}
\caption{Equilibrium phase diagrams in the $\phi_{0}$--$a_{2}$ plane. The four panels show the phase diagrams for four different choices of $a_{1}$ ($a_{1} \in \{0.01,\,0.02,\,0.025,\,0.05\}$). The solid brown curve shows the boundary that follows from the linear stability analysis (LSA), while the solid black curves provide binodals for coexistence. The black points indicate where the coexistence was obtained numerically using the grid resolution of $700$. The colour map indicates the phases belonging to~the free-energy minima forming the convex hull. Coexisting phases are shown using a hashing of~the two contributing shades. The colouring and labeling is as follows: homogeneous (white, H), charge-separated (green, S), microphase droplets (pink, D), and microphase lamella (red, L). The blue dashed vertical line indicates the cut plane for the $a_{1}$--$a_{2}$ phase diagram construction in Fig.~\ref{2-mode-dia}.}
\label{phi0-r}
\end{figure}

\paragraph{Results.} 
For $a_{1} < 0.25$, the equilibrium phase diagrams in the $\phi_{0}$--$a_{2}$ plane in Fig.~\ref{phi0-r} exhibit a critical point above which only the homogenous phase is found (Fig.~\ref{phi0-r}, white region), because of the electrostatic repulsion prevailing over the Cahn-Hilliard phase separation. By symmetry, the critical point is at~$\phi_{0} = 0.5$ for all $a_{1}$. For $a_{1} = 0.01$, the critical point is $[\phi_{0},\,a_{2}] = [0.5,\,0.0169]$ (Fig.~\ref{phi0-r}, top left) and the critical point rises to larger values of $a_{2}$ as $a_{1}$ increases. Below this critical point, there is a first-order phase separation with coexistence between H and D (Fig.~\ref{phi0-r}, hatched magenta-white regions), and D and L (Fig.~\ref{phi0-r}, hatched magenta-red regions). The lamellar phase (Fig.~\ref{phi0-r}, red region) always surrounds the centre of the phase diagram for sufficiently large $a_{2}$ above the charge-separated phase (Fig.~\ref{phi0-r}, green region). 

These features are reminiscent of the Swift-Hohenberg phase diagram~\cite{knobloch}. Coexistence lines are found to extend to the critical point upon refining the calculation mesh to 1000 grid points per coordinate, as expected for~a~2D system~\cite{knobloch}. As $a_{2}$ is reduced away from the critical point, the droplet phase first expands in $\phi_{0}$, before tapering off at even lower $a_{2}$ values. Below~this point, there is a~coexistence between the homogeneous and lamellar phase (Fig.~\ref{phi0-r}, hatched red-white regions). 
%The upper part of the $\phi_{0}$--$a_{2}$ phase diagrams (\autoref{phi0-r}) resembles that of~the two-dimensional Swift-Hohenberg model, reported in the paper by Knobloch \textit{et~al.},~\cite{knobloch}. 

There are some notable differences of our analytical phase diagram with respect to the Swift-Hohenberg phase diagram. First, the inner H/D coexistence line intersects the LSA curve, as the H/D region approaches the critical point outside of the region delimited by~the LSA. %We have also verified that this is the case in the 3- to 6-mode harmonic approximations (data not shown). We believe that the reason that the LSA curve lies outside of~the H/D coexistence region sufficiently close to the coexistence point is that in a phase-field theory microphase separation gives rise to a pattern formation regime which can break the binodal-spinodal-spinodal-binodal ordering found in,~\textit{e.g.}, a van der Waals gas. %That is, LSA constructions act on the homogeneous free-energy branch, while mode-type constructions leave that branch and move onto a~patterned branch. In addition, the LSA does not exhaustively describe the stability of the homogeneous H phase. It provides only a rough image: the stability of H against infinitely small perturbations. In the present case, it fails to~catch the boundary between H and S phases, for example. 
Second, %between the Swift-Hohenberg and charged phase separation models %Cahn-Hilliard/Debye-H\"{u}ckel pha-se diagrams 
in our model the charge-separated phase S is present for the lowest values of parameter $a_{2}$ (Fig.~\ref{phi0-r}). %In this lower-$a_{2}$ limit, the lamellar phase becomes unstable and a region of~a~charge-separated bulk phase appears, surrounded by the homogeneous phase. The lamellar phase loses stability because electrostatic interactions become negligible compared with~the Cahn-Hilliard part of the free-energy functional in the limit $a_{2} \ll 1$. The S region borders with H, H/L and L phases without coexistence with any of them. This S region is a result of the Cahn-Hilliard part of the free-energy functional. 
Third, there is a coexistence region H/L for low $a_2$. %is another new feature. % of the Cahn-Hilliard/Debye-H\"{u}ckel equilibrium phase diagram, in comparison with~its Swift-Hohenberg counterpart. 

\begin{figure}[tb]
\centering
\includegraphics[width = 0.9\linewidth]{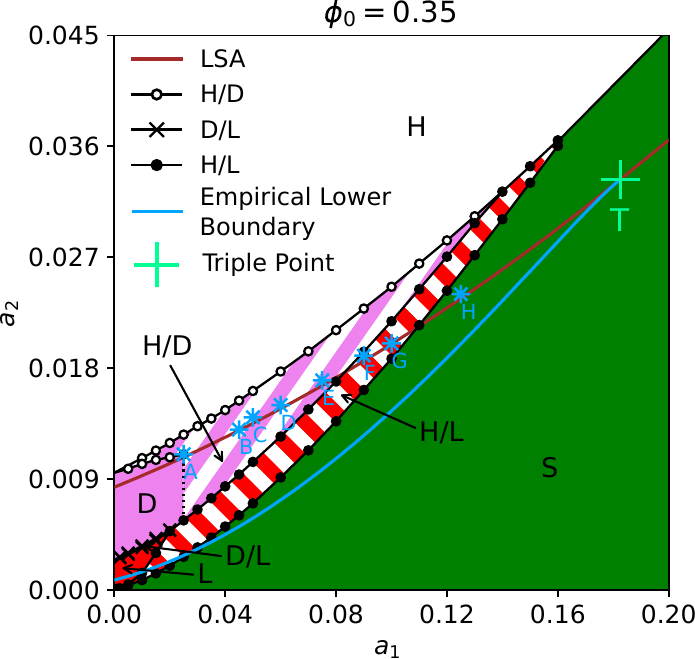}
\vspace*{-0.5cm}
\caption{Analytical equilibrium phase diagram in the $a_{1}$--$a_{2}$ plane for $\phi_{0} = 0.35$. The coloring is the same as in Fig.~\ref{phi0-r}. The solid and open dots, as well as crosses, indicate the $a_{1}$-slice values from which the diagram is constructed. The~blue asterisks indicate the points studied numerically, Fig.~\ref{fig:phase-diagram-sim}. The large green plus marked ``T'' indicates the~triple point based on the LSA. The blue curve, given by the formula $a_{2} = (23a_{1} - 2\hat{\phi} + \sqrt{\Phi})^{2}/26^{2}$, where $\Phi \equiv 4\hat{\phi}^{2} - 14\hat{\phi}a_{1} - 17a_{1}^{2}$, is the empirical lower boundary of the modulated-phase region from Fig.~\ref{fig:phase-diagram-sim}.}
\label{2-mode-dia}
\end{figure}

To directly compare to the numerical simulations shown in the main text, $\phi_{0} = 0.35$ section lines are drawn over the $\phi_{0}$--$a_{2}$ diagrams (Fig.~\ref{phi0-r}) are these are employed to construct the $a_{1}$--$a_{2}$ diagram (Fig.~\ref{2-mode-dia}). Colour maps displayed in Fig.~\ref{2-mode-dia} represent phases in thermodynamic equilibrium as predicted by the two-mode harmonic approximation. Below~$a_{1}=0.020$ a~full sequence of~phases and coexistences is observed. Between~$a_{1}=0.020$ and $a_{1}=0.025$, the pure L phase is absent. Above $a_{1}=0.025$, only H, H/D, H/L and S phases are stable. In~addition, above~$a_{1} = 0.13$, there are only H, H/L and S stable phases. Finally, above~$a_{1} = 0.16$, \emph{no} microstructured pattern is stabilized (Fig.~\ref{2-mode-dia}). 

%Increasing $a_{1}$ above the threshold value of $0.25$ causes all microseparated phases to~disappear. This justifies the prediction of the triple point $T = [\Gamma,\,\Gamma^{2}]$ within the linear stability analysis (LSA) with $\Gamma \equiv 1/4 - 3(\phi_{0} - 1/2)^{2}$. For $a_{1} > \Gamma$, there are no stable microphase separated patterns and the homogeneous solution H coexists with the charge-separated phase S along~the parabola $a_{2} = \phi_{0}(1 - \phi_{0})a_{1}$. The coexistence of the phase S with any other phase has not been observed in any of the predicted phase diagrams. This may be because the total charge separation S occurs in the region of~the phase diagram where electrostatics becomes negligible. 

The $a_{1}$--$a_{2}$ diagram (Fig.~\ref{2-mode-dia}) can be compared to~the results obtained numerically (Fig.~\ref{fig:phase-diagram-sim} in the main text) to gain insight into our previous observations. In Fig.~\ref{fig:phase-diagram-sim}, we focus on 8 particular points, labelled A-H, and showed the long-time limit snapshots (Fig.~\ref{fig:equilibrium-states}). We can now see that patterns A, B, and C indeed fall into the H/D coexistence region. The pattern in panel A is at the border of the pure droplet-suspension domain, which makes sense given that droplets cover the simulation domain. Pattern D %is more difficult to explain, as it appears to have 
has homogeneous, lamellar, and droplet features. %, though the ``lamellar'' aspects could be classified as elongated droplets. 
An extremely thin region of coexistence between droplet and lamellae is in fact predicted from the analytical model (Fig.~\ref{2-mode-dia}b, D/L region at the bottom left corner), although we note pattern D is relatively far from this region. Points E, F, and G are labyrinthine patterns that are either very close to (E) or inside (F and G) the H/L coexistence region. %Points H-K show droplets of various size, with~different amounts of polydispersity. Of course, within the analytical model we cannot reproduce by construction amorphous phases, so that the 2-mode approximation cannot be used to decide whether they are thermodynamically stable as our numerics suggest or whether they are all long-lived metastable states. %This leads us to suspect that these simulations are on their way to the fully charge-separated phase S, but have not yet reached this phase within~the simulation time considered. This trend is most pronounced for points H and I, which have relatively high $a_{1}$ values. \\
More in general, amorphous patterns tend to be observed numerically in regions where our two-mode approximation suggests coexistence.

\newpage

\clearpage

% Reset the figure counter to 0
\setcounter{figure}{0}

% Redefine the figure number format to S1, S2, etc.
\renewcommand{\thefigure}{S\arabic{figure}}

%Reset the equation counter to 0
\setcounter{equation}{0}

% Redefine the equation number format to S1, S2, etc.
\renewcommand{\theequation}{S\arabic{equation}}

\section*{Supplemental Material: Arrested phase separation and amorphous patterns in charged binary fluids}

\subsection{Additional numerical and analytical results}

Fig.~\ref{fig:equilibrium-states-suppl} shows all points A-H in the phase diagram for $\phi_0=0.35$ (Fig.~\ref{fig:phase-diagram-sim} in the main text), together with the two-dimensional radial distribution function and the two-dimensional structure factor.

Fig.~\ref{fig:extraconfigs_free_energy_vs_a2} shows all initial conditions and final steady state (metastable) patterns obtained with the large scale simulations for which we computed the average free energy density in Fig.~3 of the main text. The hexagonal islands in uniform background initial conditions leads to surprisingly complex patterns. %, qualitatively similar to quasicrystals. 
That configuration and the lamellar stripe configuration are examples of configurations with homogeneous-droplet and homogeneous-lamellae coexistence, predicted to be thermodynamically stable by our two-mode analysis (see End Matter). We also note that for the timescale corresponding to the final patterns in Fig.~3 of the main text or Fig.~\ref{fig:extraconfigs_free_energy_vs_a2}, the average free energy densities are still very slowly decreasing but the magnitude of the change is much smaller than the difference in average free energy densities between patterns, so the change is not sufficient to affect the ranking between structures. Structure factors corresponding to these patterns are shown in Fig.~\ref{fig:extraconfigs_free_energy_vs_a2_sk}. These confirm the amorphous patterns of the labyrinthine patterns and of the metastable steady state reached starting from regular lamellae for $a_2=0.021$ (top row, rightmost plot in Figs.~\ref{fig:extraconfigs_free_energy_vs_a2} and~\ref{fig:extraconfigs_free_energy_vs_a2_sk}).

An additional homogeneous-initiated state diagram for $\phi_0=0.4$ is shown in Fig.~\ref{fig:pd0.4}, with some (in general metastable) steady states shown in Fig.~\ref{fig:eqst0.4}. The results show that amorphous behaviour is not limited to the parameters shown in the main text, but a robust feature of the system. This numerical state diagram also compares well with its analytical equilibrium counterpart constructed in the two-mode harmonic approximation, displayed in Fig.~\ref{fig:0.4-anal}.  
%Hexagonal-droplet pattern in panel A lies on the borderline of the pure hexagonal-droplet region. Point B indeed falls into the H/D coexistence region. Finally, points C and D exhibiting labyrinthine patterns indeed belong to the H/L coexistence region, as expected.  %$\phi_0=0.35$. 

\subsection{Additional numerical details}

%\subsection{Direct numerical simulations}

The equations of motion, Eqs.~\eqref{eq:non-dim-phi-psi}, are solved numerically using either an in-house spectral code (Figs.~\ref{fig:phase-diagram-sim},~\ref{fig:equilibrium-states},~\ref{fig:pd0.4} and~\ref{fig:eqst0.4}), or an MPI-parallel fully-dealiased pseudo-spectral code developed within the Dedalus framework~\cite{Burns2020} (Figs.~\ref{fig:free_energy_vs_a2} and~\ref{fig:fig4}). 

\begin{figure*}[!h]
    \centering
    \includegraphics[width=1.\linewidth]{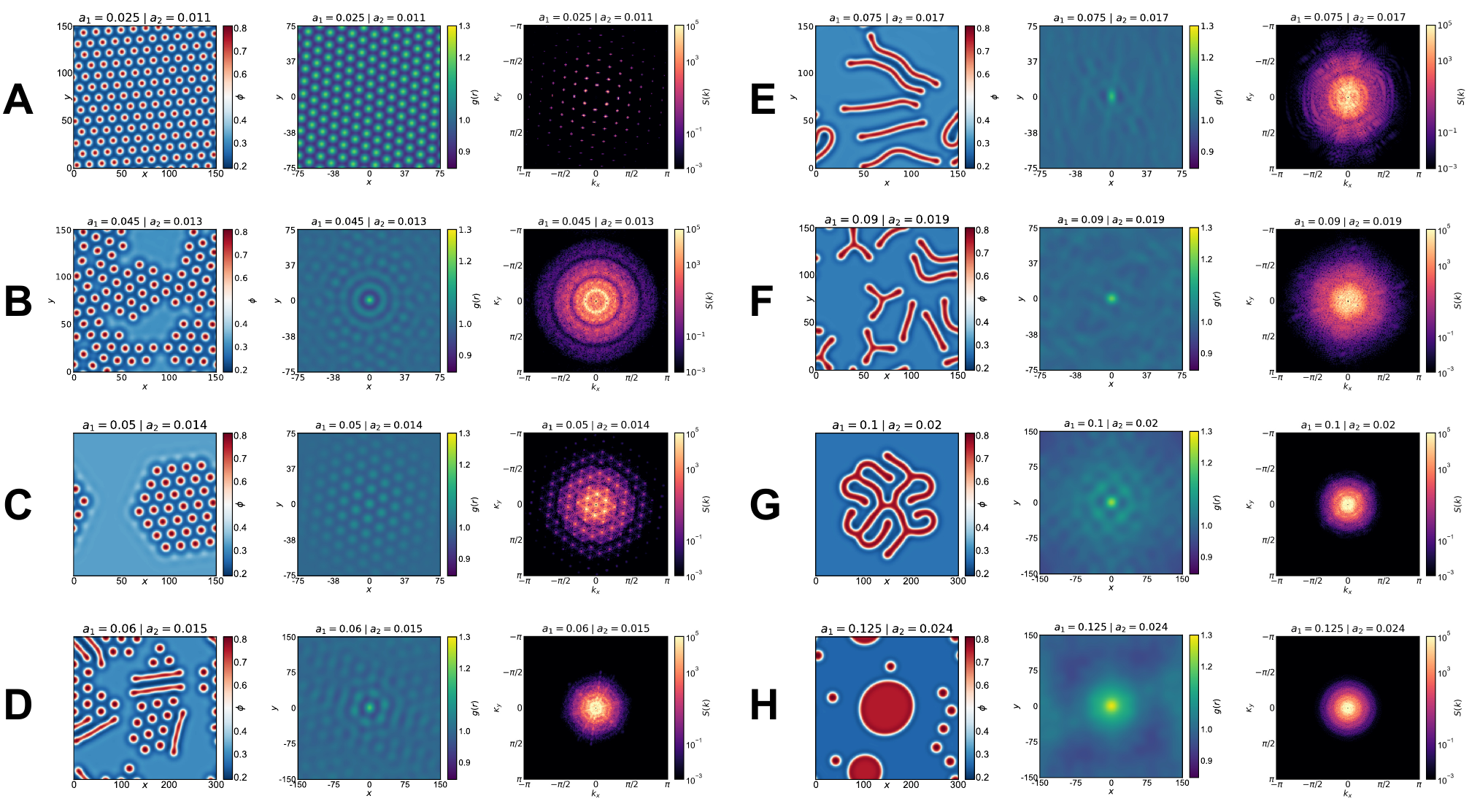}
    \caption{Snapshots (left column), two-dimensional radial distribution function (middle column) and two-dimensional structure factor (right column, logarithmic scale for the colour bar) corresponding to the points A-H in Fig.~\ref{fig:phase-diagram-sim}.}
    \label{fig:equilibrium-states-suppl}
\end{figure*}

\begin{figure*}[!h]
    \centering
\includegraphics[width=1.\linewidth]{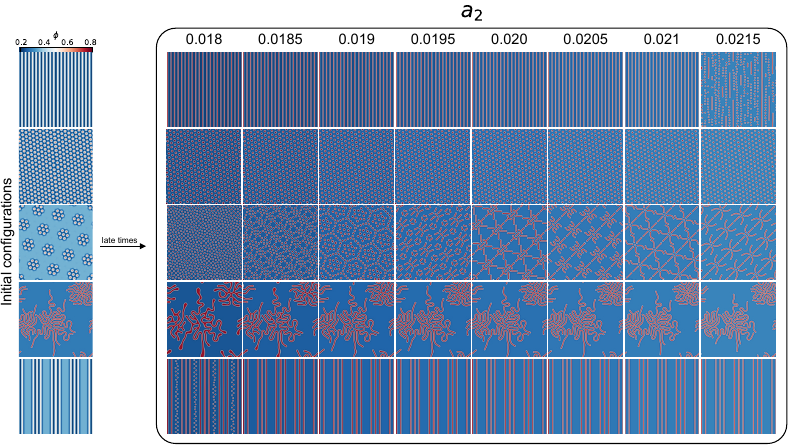}
    \caption{Initial and late time configurations for all configurations corresponding to the curves shown in Fig.~\ref{fig:free_energy_vs_a2}.}
    \label{fig:extraconfigs_free_energy_vs_a2}
\end{figure*}

\begin{figure*}[!h]
    \centering
\includegraphics[width=1.\linewidth]{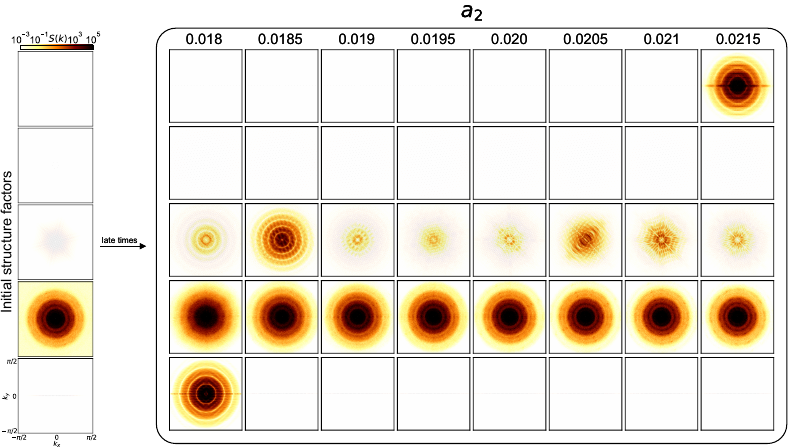}
    \caption{Initial and late time structure factors for all configurations shown in Fig.~\ref{fig:extraconfigs_free_energy_vs_a2}. As the logarithmic scale is the same in all plots, for regular patterns peaks may require magnifications to be seen.}
    \label{fig:extraconfigs_free_energy_vs_a2_sk}
\end{figure*}

\begin{figure}[!t]
    \centering
    \includegraphics[width=0.9\linewidth]{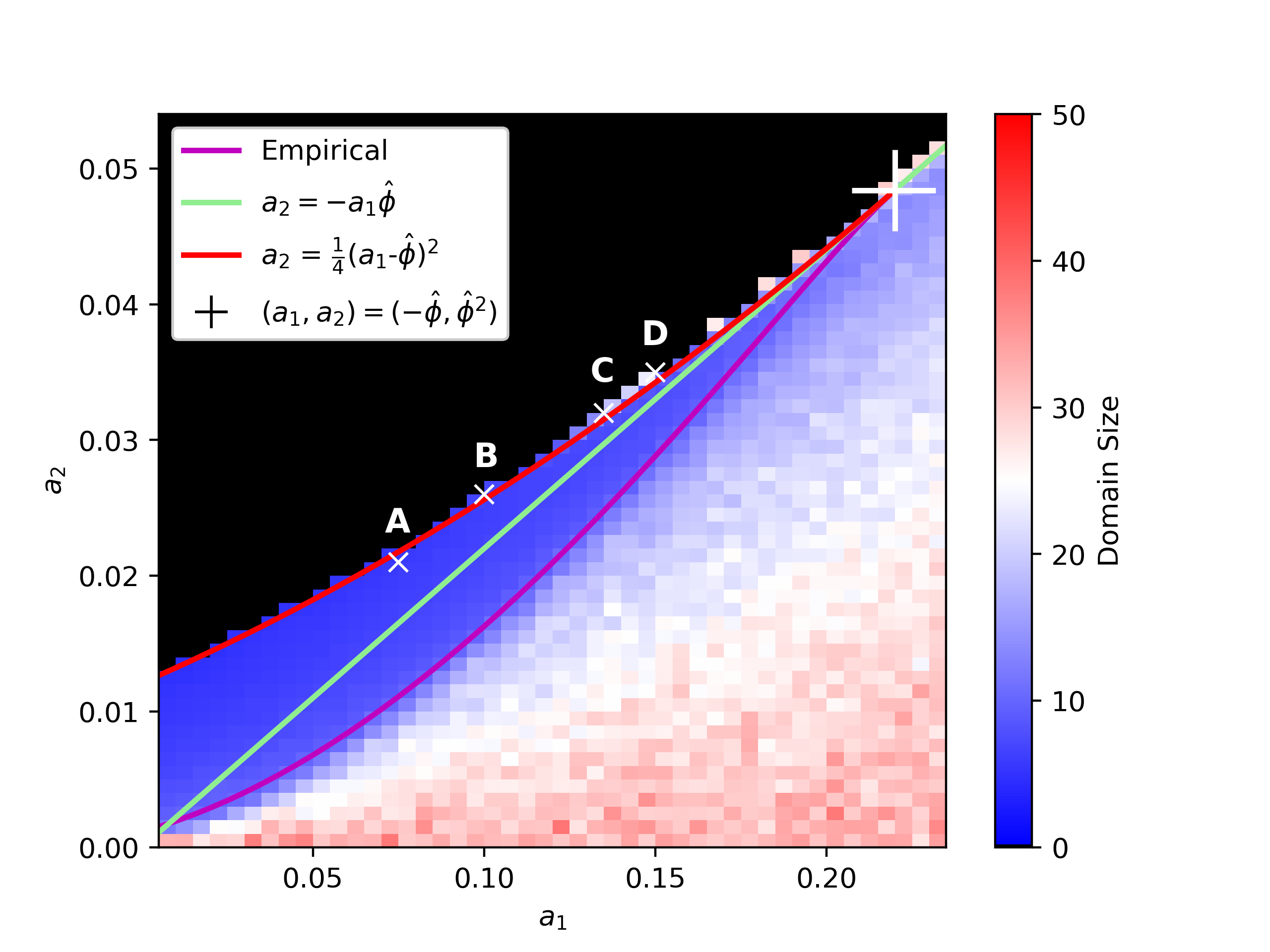}
    \caption{State diagram for $\phi_0=0.4$ with the initial condition of homogeneous and uniform noise of $\delta\phi=0.1$, with theoretical phase boundaries from linear stability analysis overlaid (red line: upper boundary, cyan line: lower boundary). A numerical estimate (pink) for the lower boundary is also shown.  Each pixel represents the domain size of a state close to equilibrium. %- this results in the three expected regions. 
Steady states for A-D are shown in Fig.~\ref{fig:eqst0.4}.}
    \label{fig:pd0.4}
\end{figure}

\begin{figure}[!t]
    \centering
    \includegraphics[width=1\linewidth]{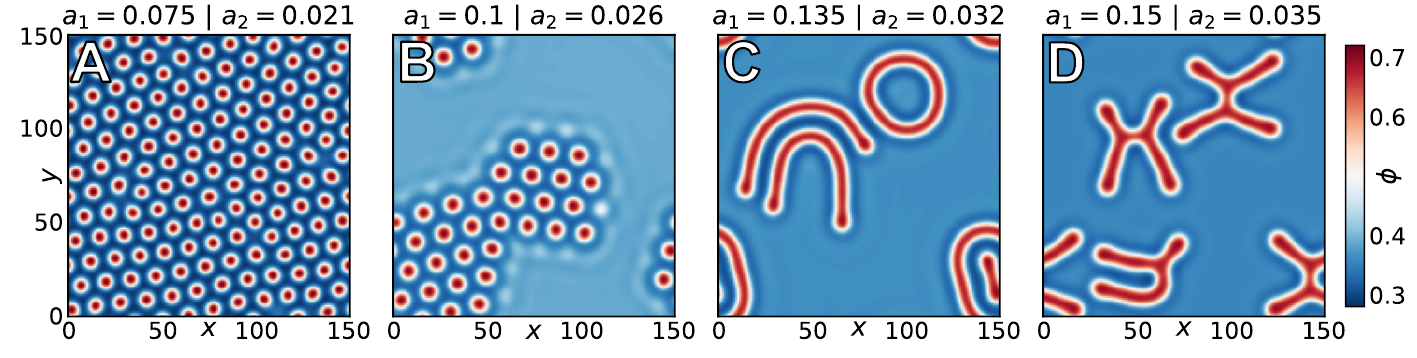}
    \caption{Steady states for points A-D in Fig.~\ref{fig:pd0.4}.}
    \label{fig:eqst0.4}
\end{figure}

\begin{figure}[!t]
    \centering
    \includegraphics[width=0.9\linewidth]{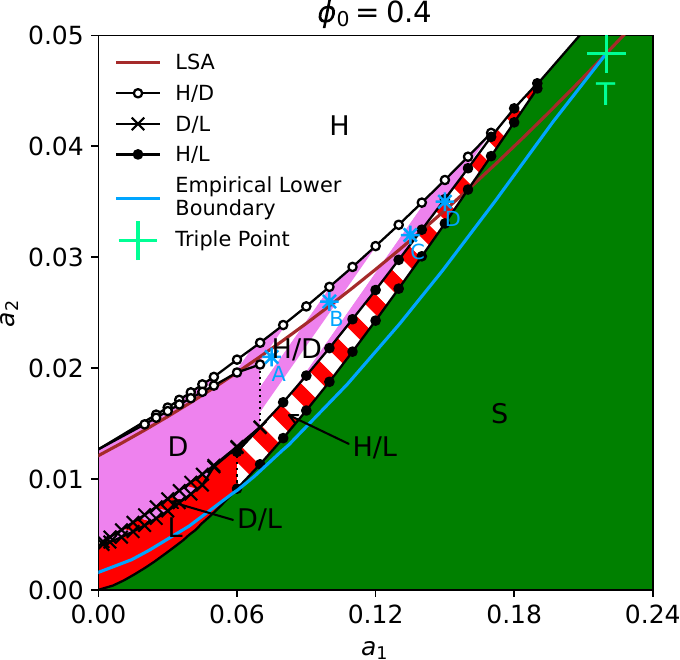}
    \caption{Analytical equilibrium phase diagram in the $a_{1}$-$a_{2}$ plane, constructed in the two-mode harmonic approximation for $\phi_{0} = 0.4$, to be compared with its numerical counterpart in Figs.~\ref{fig:pd0.4},~\ref{fig:eqst0.4}. The colour code is the same as in Figs.~\ref{phi0-r},~\ref{2-mode-dia}.}
    \label{fig:0.4-anal}
\end{figure}

For the in-house code, in simulation units, the configurations in Fig.~\ref{fig:equilibrium-states}A,C and F, in Fig.~\ref{fig:equilibrium-states-suppl}A,B,C,E,F and in Fig.~\ref{fig:eqst0.4} correspond to systems of $150\times 150$ whilst the configurations in Fig.~\ref{fig:equilibrium-states}D,H and G  and in Fig.~\ref{fig:equilibrium-states-suppl}D,H and G correspond to systems of $300\times 300$ (sizes are given in dimensionless units).

For the Dedalus simulations, time-discretisation is performed using a fourth-order semi-implicit backward differentiation formula (BDF) scheme~\citep{Wang2008} with the timestep ranging from $0.1$ to $0.001$.
The $\phi$ and $\psi$ fields are represented through a spectral decomposition based on Fourier modes; as mentioned in the main text, we consider 2D and 3D simulation domains of size $L^2$ and $L^3$, respectively. The spectral resolution is set by specifying the number of modes, $N$, which was set to be the same in each direction. 
Production runs are carried out using $N=1024$ with $L=1000$, in 2D, while in 3D we set $N=512$ with $L=800$ for the runs in Fig.~\ref{fig:fig4}A-C, and $N=128$ with $L=200$ for those in Fig.~\ref{fig:fig4}D. All simulations are evolved until they reached statistically steady-state.  A typical production run starts from a random initial condition and is performed on 256 CPU cores for approximately 48 hours. 

For the characterisation of the 3D structures in Fig.~\ref{fig:fig4}, the local curvature, Euler characteristic and average domain size are computed. The domain size $L(t)$ is the inverse of the first moment of the structure factor
\begin{equation}
    L(t)=2\pi \frac{\int S(k)dk}{\int kS(k)dk},
\end{equation}
where $S(k)$ is the structure factor. The Euler characteristic $\chi$ is a topological invariant, which is given by
\begin{equation}
    \chi = V-E+F,
    \label{euler}
\end{equation}
for two dimensional surfaces, where $V$ is the number of vertices, $E$ edges and $F$ faces. Notably, $\chi=0$ for a torus and $\chi=2$ for a sphere. The Euler characteristic is additive and so arrested states composed of $N$ structures topological equivalent to a sphere will have $\chi=2N$. The Euler characteristic is computed directly through Eq.~\eqref{euler} on the triangulated isosurfaces $\phi\approx \phi_m$, where $\phi_m$ is the midpoint between the two dominant peaks of the distribution of $\phi$. The triangulation was performed using the marching cubes algorithm~\cite{lewiner2003} through the implementation of \texttt{skimage.measure.marching\_cubes}. The local curvature and visualisation of the 3D structures were done using the software Paraview~\cite{ahrens2005}. 

\subsection*{Mapping of free energy differences to physical units}

In Fig.~\ref{fig:free_energy_vs_a2} of the main text, free energy densities and their differences are measured in dimensionless units. The dimensionless form for the free energy density is
\begin{equation}\label{nondimfreeenergydensity}
    f=\frac{1}{4}\phi^2(\phi-1)^2+\frac{1}{2}\left(\nabla\phi\right)^2-\frac{1}{2a_2}\left(\nabla\psi\right)^2-\frac{a_1}{2a_2}\psi^2+\phi\psi,
\end{equation}
where $\psi$ is also dimensionless as described in the text. 

To map free energy differences to physical values, we observe that, in Eq.~(\ref{nondimfreeenergydensity}), $f$ is measured in units of $a$. An estimate of $a$ is given by $\gamma/\delta$, with $\gamma$ a typical value of the surface tension between the two fluids, for instance $\sim 1$ mN/m, and $\delta$ a typical interfacial thickness, for instance $1$ nm. We therefore get that a physically reasonable value for $a$ is $a \sim 10^6$ Pa, hence a difference of $10^{-6}$ corresponds to $1$ Pa. To map to a free energy difference, we note that the typical lengthscale over which lamellae branch or bend in labyrinthine states is similar to the pattern lengthscale. This is the relevant scale as the entropy of an amorphous state of this size can be estimated as $k_BT$ (in the same way in which the entropy of a polymer can be estimated as $k_BT$ times the number of monomers). If we consider patterns of size $\sim 100$ nm, which can be relevant for polymeric fluids, then we obtain that a difference of $1$ $k_BT$ in free energy corresponds to a free energy density difference of $\sim 4.1 \times 10^{-6}$ in dimensionless units. We therefore expect that if an amorphous pattern is within this range of an ordered one, it will in practice likely outcompete it due to entropic effects. In eight of the nine parameter sets studied in Fig.~\ref{fig:free_energy_vs_a2} the difference between the minimum and the most stable amorphous state we considered was below $k_BT$ in this mapping.

\subsection*{Supplementary Movie Captions}

{\bf Suppl. Movie 1.} Dynamics corresponding to the formation of crystals of droplets in 2D for $a_1=0.05$, $a_2=0.01375$. The system size if $1000\times 1000$. Left and right panels correspond to $\phi$ and $\psi$ respectively.

{\bf Suppl. Movie 2.} Dynamics corresponding to the formation of an amorphous labyrinthine state for $a_1=0.1$, $a_2=0.0208$. The system size if $1000\times 1000$. Left and right panels correspond to $\phi$ and $\psi$ respectively.

\clearpage

\begin{thebibliography}{99}
\bibitem{seul1995} M. Seul and D. Andelman, %Domain shapes and patterns: the phenomenology of modulated phases.
{\it Science} {\bf 267}, 476 (1995). %476-493
\bibitem{andelman2009} D. Andelman and R.~E. Rosensweig, 
%Modulated phases: Review and recent results. 
{\it J. Chem. Phys. B} {\bf 113}, 3785 (2009). %3785--3798 
\bibitem{jackson1994} D.~P. Jackson, R.~E. Goldstein, and A.~O. Cebers, %Hydrodynamics of fingering instabilities in dipolar fluids. 
{\it Phys. Rev. E} {\bf 50}, 298 (1994).
\bibitem{haw2010} M.~D. Haw, %Growth kinetics of colloidal chains and labyrinths
{\it Phys. Rev. E} {\bf 81}, 031402 (2010).
\bibitem{low1994} U. L\"ow, V.~J. Emery, K. Fabricius, and S.~A. Kivelson,
%Study of an Ising model with competing long- and short-range interactions.
{\it Phys. Rev. Lett.} {\bf 72}, 1918 (1994). %1918-1921
\bibitem{schmalian2000} J. Schmalian and P.~G. Wolynes,
%Stripe glasses: self-generated randomness in a uniformly frustrated system.
{\it Phys. Rev. Lett.} {\bf 85}, 836 (2000). %836-839
\bibitem{liu2019} Y. Liu and Y. Xi, %Colloidal systems with a short-range attraction and long-range repulsion: Phase diagrams, structures, and dynamics. 
{\it Curr. Opin. Colloid Interface Sci.} {\bf 39}, 123 (2019). %123-136 
\bibitem{spivak2004} B. Spivak and S.~A. Kivelson,
%Phases intermediate between a two-dimensional electron liquid and Wigner crystal.
{\it Phys. Rev. B} {\bf 70}, 155114 (2004).
\bibitem{stradner2004} A. Stradner, H. Sedgwick, F. Cardinaus, W.~C.~K. Poon, S.~U. Egelhaaf, and P. Schurtenberger, {\it Nature} {\bf 432}, 492 (2004). %492-495
\bibitem{campbell2005} A.~I. Campbell, V.~J. Anderson, J.~S. van Duijneveld, and P. Bartlett, %Dynamical arrest in attractive colloids: the effect of long-range repulsion, 
{\it Phys. Rev. Lett.} {\bf 94}, 208301 (2005).
\bibitem{chakrabarti2011} S. Chakrabarti and Z. Nussinov, 
%Modulation and correlation lengths in systems with competing interactions,
{\it Phys. Rev. B} {\bf 84}, 144402 (2011).
\bibitem{ortix2006} C. Ortix, J. Lorenzana, and C. Di Castro, 
%Frustrated phase separation in two-dimensional charged systems
{\it Phys. Rev. B} {\bf 73}, 245117 (2006).
\bibitem{vannimenus1977} J. Vannimenus and G. Toulouse, %Theory of the frustration effect. II. Ising spins on a square lattice. 
{\it J. Phys. C.} {\bf 10}, L537 (1977).
\bibitem{zaccarelli2007} E. Zaccarelli, %Colloidal gels: equilibrium and non-equilibrium routes,
{\it J. Phys.: Condens. Matt.} {\bf 19}, 323101 (2007).
\bibitem{tarjus2005} G. Tarjus, S.~A. Kivelson,
Z. Nussinov, and P. Viot,
%The frustration-based approach of supercooled liquids and the glass transition.
{\it J. Phys.: Condens. Matter} {\bf 17}, R1143 (2005). %R1143-R1182
\bibitem{sciortino2004} F. Sciortino, S, Mossa, E, Zaccarelli, and P. Tartaglia,
{\it Phys. Rev. Lett.} {\bf 93}, 055701 (2004).
\bibitem{kivelson1995} D. Kivelson, S.~A. Kivelson,
X.~L. Zhao, Z. Nussinov, and G. Tarjus,
%A thermodynamic theory of supercooled liquids.
{\it Physica A} {\bf 219}, 27 (1995). %27-38
\bibitem{boroudjerdi2005} H. Boroudjerdi, Y.~W. Kim, A. Naji, X. Netz, X. Schlagberger, and R.~R. Netz, {\it Phys. Rep.} {\bf 416}, 129 (2005).
\bibitem{dean2014} D.~S. Dean, J. Dobnikar, A. Naji, and R. Podgornik (Eds.), {\it Electrostatics of Soft and Disordered Matter} (CRC Press, Boca Raton, 2014).
\bibitem{zheng2021} B. Zheng, Y. Avni, D. Andelman, and R. Podgornik, {\it J. Phys. Chem. B} {\bf 125}, 7855 (2021).
\bibitem{shin2017} Y. Shin and C.~P. Brangwynne, 
%Liquid phase condensation in cell physiology and disease
{\it Science} {\bf 357}, eaaf4382 (2017).
\bibitem{brazovskii1975} S.~A. Brazovskii,
%Phase transition of an isotropic system to a nonuniform state. 
{\it Sov. J. Exp. Theor. Phys.} {\bf 41}, 85 (1975).
\bibitem{brazovskii1987} S.A. Brazovskii,I.E. Dzyaloshinskii,andA.R. Muratov,
%Theory of weak crystallisztion,
{\it Zh. Eksp. Teor. Fiz.} {\bf 93}, 1110 (1987). %1110-1124.
\bibitem{elder2002} K.~R. Elder, M. Katakowski, M. Haataja and M. Grant, 
%Modeling Elasticity in Crystal Growth, 
{\it Phys. Rev. Lett.} {\bf 88}, 245701 (2002).
\bibitem{lowe2011} M. Lowe, 
%Structural organization of the Golgi apparatus. 
{\it Curr. Opin. Cell Biol.} {\bf 23}, 85 (2011). %85-93.
\bibitem{SI} See Supplemental Materials for additional numerical results and numerical details, and for Supplementary Movies.
\bibitem{bergeon2008} A. Bergeon, J. Burke, E. Knobloch, and I. Mercader, %Eckhaus instability and homoclinic snaking. 
{\it Phys. Rev. E} {\bf 78}, 046201 (2008).
\bibitem{kozyreff2006} G. Kozyreff and S. J. Chapman, {\it Phys. Rev. Lett.} {\bf 97}, 044502 (2006).
\bibitem{henrich2010} O. Henrich, K. Stratford, D. Marenduzzo and M.~E. Cates, 
%Ordering dynamics of blue phases entails kinetic stabilization of amorphous networks, 
{\it Proc. Natl. Acad. Sci. USA} {\bf 107}, 13212 (2010). %13212–13215
\bibitem{chaikin1995} P.~M. Chaikin and T.~C. Lubensky, {\it Principles of condensed matter physics}, Cambridge University Press, Cambridge (1995).
\bibitem{henrich2011} O. Henrich, K. Stratford, M.~E. Cates, and  D. Marenduzzo, %Structure of blue phase III of cholesteric liquid crystals. 
{\it Phys. Rev. Lett.} {\bf 106}, 107801 (2011).
\bibitem{wright1989} D.~C. Wright and N.~D. Mermin,
%Crystalline liquids: the blue phases.
{\it Rev. Mod. Phys.} {\bf 61}, 385 (1989). %385-432
\bibitem{carenza2022} L.~N. Carenza, G. Gonnella, D. Marenduzzo, G. Negro and E. Orlandini,
%Cholesteric Shells: Two-Dimensional Blue Fog and Finite Quasicrystals
{\it Phys. Rev. Lett.} {\bf 128}, 027801 (2022).
\bibitem{capuani2004} F. Capuani, I. Pagonabarraga and D. Frenkel, 
%Discrete solution of the electrokinetic equations. 
{\it J. Chem. Phys.} {\bf 121}, 973 (2004) %973-986
\bibitem{giupponi2011} G. Giupponi and I. Pagonabarraga 
%Colloid electrophoresis for strong and weak ion diffusivity. 
{\it Phys. Rev. Lett.} {\bf 106}, 248304 (2011).
\bibitem{jiang2019} H. Jiang, P.~G. Debenedetti and A.~Z. Panagiotopoulos, 
%Nucleation in aqueous NaCl solutions shifts from 1-step to 2-step mechanism on crossing the spinodal, 
{\it J. Chem. Phys.} {\bf 150}, 124502 (2019).
\bibitem{sing2020} C.~E. Sing and S.~L. Perry, 
%Recent progress in the science of complex coacervation. 
{\it Soft Matter} {\bf 16}, 2885 (2020).
\bibitem{dutagaci2021} 
B. Dutagaci, G. Nawrocki, J. Goodluck, A.~A. Ashkarran, C.~G. Hoogstraten, L.~J. Lapidus, and M. Feig, 
%Charge-driven condensation of RNA and proteins suggests broad role of phase separation in cytoplasmic environments
{\it eLife} {10}, e64004 (2021).
\bibitem{brackley2017} C.~A. Brackley, B. Liebchen, D. Michieletto, F. Mouvet, P.~R. Cook and D. Marenduzzo, 
%Ephemeral protein binding to DNA shapes stable nuclear bodies and chromatin domains. 
{\it Biophys. J.} {\bf 112}, 1085 (2017). %1085-1093
\bibitem{chull}
O. Ryll, S. Blagov, and H. Hasse,
%Convex envelope method for~the determination of~fluid phase diagrams}",
{\it Fluid Phase Equilibria} {\bf 324}, 108 (2012).
\bibitem{2mode} M. J. Robbins, A.~J. Archer, U. Thiele, and E. Knobloch,
%Modeling the structure of~liquids and crystals using one- and two-component modified phase-field crystal models
{\it Phys. Rev. E} {\bf 85}, 061408 (2012).
\bibitem{knobloch} U. Thiele, A.~J. Archer, M.~J. Robbins, H. Gomez and E. Knobloch, %Localized states in~the conserved Swift-Hohenberg equation with~cubic nonlinearity}",
{\it Phys. Rev. E} {\bf 87}, 042915 (2013).
\bibitem{chdh} E. Posp\'{i}\v{s}ilov\'{a}, L.~H. Grant, D. Marenduzzo, T.~N. Shendruk, and J. de~Graaf, %Equilibrium Phase Diagram of~a~Two-Dimensional Phase-Field Model with~Screened Coulomb Interaction}",
in preparation. %{\it Phys. Rev. E} (2027).
\bibitem{lewiner2003} T. Lewiner, H. Lopes, A.~W. Vieira, and G. Tavares, %Efficient Implementation of Marching Cubes' Cases with Topological Guarantees.
{\it J. Graph. Tools} {\bf 8}, 1 (2003). %1-15
\bibitem{ahrens2005} J. Ahrens, B. Geveci, and C. Law, %ParaView: An End-User Tool for Large Data Visualization. 
{\it Visualization Handbook}, Elsevier (2005). %ISBN-13: 9780123875822
\bibitem{Burns2020} K.~J. {Burns}, G.~M. {Vasil}, J.~S. {Oishi}, D. {Lecoanet}, and B.~P. {Brown}, {\it Phys. Rev. Res.} {\bf 2}, 023068 (2020).
\bibitem{Wang2008} D. Wang and S.~J. Ruuth, {\it J. Comput. Math.} {\bf 26}, 838 (2008).
%extra refences to possibly add
%Carretas-Talamante, A. G., López, J. B. Z., Lázaro-Lázaro, E., Elizondo-Aguilera, L. F., & Medina-Noyola, M. (2023). Non-equilibrium view of the amorphous solidification of liquids with competing interactions. The Journal of Chemical Physics, 158(6), 064506. https://doi.org/10.1063/5.0132525
%\bibitem{henrich2017} O. Henrich and D. Marenduzzo, 
%The secret of the blue fog. 
%{\it Phys. World} {\bf 30}, 25 (2017). %25-29
%\bibitem{liebchen2015} B. Liebchen, D. Marenduzzo, I. Pagonabarraga and M.~E. Cates, %Clustering and pattern formation in chemorepulsive active colloids. 
%{\it Phys. Rev. Lett.} {\bf 115}, 258301 (2015).
%\bibitem{besseling2026} 
%T.~H. Besseling, B. van der Meer, B. Liu, L. Filion, A. Imhof and A. van Blaaderen,
%%An equilibrium rotator glass-forming phase for long-ranged repulsive colloidal rods
%{\it Nat. Commun.} {\bf 17}, 2410 (2026). 
\end{thebibliography}
\end{document}